\documentclass[paperpaper,journal]{IEEEtran}  % 修正文档类选项
\usepackage{amsmath,amsfonts}
\usepackage{algorithmic}
\usepackage{algorithm}
\usepackage{array}
\usepackage{textcomp}
\usepackage{stfloats}
\usepackage{url}
\usepackage{verbatim}
\usepackage{graphicx}
\usepackage{cite}
\usepackage{amssymb}

\usepackage{amsthm}

\usepackage{xcolor}

\usepackage[caption=false,font=footnotesize]{subfig}
\usepackage{hyperref}            % hyperref放在最后
\begin{document}

\title{Grid-Refinement SBL for AFDM Channel Estimation Under Fractional Delay and Fractional Doppler Channels}

 \author{
  Qu Luo,~\IEEEmembership{Member,~IEEE,}
  Limin Liu, 
  Zhe Li, 
  Qihao Peng,~\IEEEmembership{Member,~IEEE,} \\ 
  Haowei Wu,  ~\IEEEmembership{Member,~IEEE,} 
  Pei Xiao, ~\IEEEmembership{Senior Member,~IEEE} 
\vspace{-1em}

 \thanks{Qu Luo, Qihao Peng and Pei Xiao are with the 5G \& 6G Innovation Centre, University of Surrey, U.K. (email: \{q.u.luo, q.peng, p.xiao\}@surrey.ac.uk). }
 \thanks{ Limin Liu and Haowei Wu are with the School of Microelectronics and Communication Engineering, Chongqing University, China. (email: liulimin4858@163.com, wuhaowei@cqu.edu.cn).}
  \thanks{Zhe Li is with the Research and Development Department, China Academy of Launch Vehicle Technology, Beijing, China. (email: zheli@163.com). Corresponding author (Haowei Wu).}% <-this % stops a space
}
 
  \maketitle

\begin{abstract}
This paper investigates channel estimation for affine frequency division multiplexing (AFDM) systems over channels with fractional delay and fractional Doppler (FDFD) shifts. An exact FDFD channel model is first established, and the corresponding input--output relationship is also derived. To address the FDFD-induced delay--Doppler coupling, a first-order linear approximation grid-refinement sparse Bayesian learning (FLA-GR-SBL) method is developed. To further remedy the convergence and stability issues caused by the limited accuracy of the first-order approximation, we propose a two-dimensional grid-approximation SBL (2D-GA-SBL) algorithm, which iteratively refines the delay--Doppler grid and thereby characterizes the channel more accurately. Simulation results demonstrate that the proposed algorithms significantly outperform existing on-grid and off-grid SBL benchmarks in terms of both normalized mean square error (NMSE) and bit error rate (BER).

\end{abstract}

\begin{IEEEkeywords}
Affine frequency division multiplexing (AFDM), fractional delay and Doppler, sparse Bayesian learning.
\end{IEEEkeywords}
\vspace{-1em}
\section{Introduction}

\IEEEPARstart{A}{ffine} frequency division multiplexing (AFDM) has recently 
emerged as a promising waveform for next-generation high-mobility scenarios, 
such as vehicular and low-Earth-orbit (LEO) satellite networks \cite{ref1,zeiping,ref1.5}, 
owing to its strong robustness against Doppler spread and its enhanced 
diversity gain over doubly dispersive channels \cite{ref3,refQu}. Accurate channel 
estimation is fundamental to realizing the promised error-rate performance 
of AFDM, especially in highly dynamic environments. Existing works on AFDM channel estimation generally assume integer normalized delay and Doppler shifts for each path \cite{ref3.5,ref4,ref5,ref6}. In practice, however, both parameters are typically fractional. Such fractional delay and fractional Doppler (FDFD) components spread the channel energy across the discrete affine Fourier transform (DAFT) domain, destroying the ideal channel sparsity and severely degrading most existing estimators, including linear and diagonal-reconstruction-based schemes \cite{ref4,ref5}.

{Although extensive research has been conducted on channel estimation with integer delays, studies on FDFD channel estimation and the exact DAFT-domain input--output relation of AFDM remain limited.} In \cite{ref8}, a matched-filtering-based approach was proposed for FDFD channel estimation, in which the parameters of individual paths are estimated sequentially. However, this approach relies on the separability and approximate orthogonality of the path responses in the DAFT domain, which may not hold under FDFD channels. In \cite{ref9}, a tensor-decomposition-based method was developed, but it does not provide a unified formulation of the FDFD channel. In \cite{ref12}, SBL was adopted for efficient AFDM channel estimation, yet FDFD effects were neglected. {In addition, the authors in \cite{refYaoge} proposed a data-aided grid-evolution SBL scheme for channel estimation in AFDM-based integrated sensing and communication (ISAC) systems. While these SBL-based approaches show promise for efficient channel estimation, accurate estimation under FDFD channels calls for a framework that explicitly accounts for the coupled delay--Doppler structure of AFDM and mitigates the modeling errors associated with grid-refinement  and   approximation.}

Motivated by the above observations, this paper   investigates   FDFD channel estimation for AFDM systems. A unified FDFD channel model is first derived, and the corresponding DAFT-domain input--output relationship is obtained together with its structural properties. On this basis, a first-order linear approximation grid-refinement  SBL (FLA-GR-SBL) estimator is developed and tailored to the delay--Doppler coupling of AFDM to mitigate off-grid estimation errors. To further overcome the modeling error of the first-order approximation under severe fractional shifts, a two-dimensional grid-approximation SBL (2D-GA-SBL) algorithm is proposed, which iteratively refines the delay--Doppler grid and thereby improves both accuracy and robustness. Simulation results verify the superiority of the proposed algorithms over existing SBL-based benchmarks in terms of normalized mean square error (NMSE) and bit error rate (BER) performance.

\vspace{-1em}
\section{System Model}

\subsection{AFDM over FDFD Channels}

Let $\mathbf{x}\in {{\mathbb{A}}^{N\times 1}}$ denote a vector of $N$ quadrature amplitude modulation (QAM) symbols in the DAFT domain, where $\mathbb{A}$ is the modulation alphabet. The vector $\mathbf{x}$ is   {transformed to the time domain} via an $N$-point inverse DAFT, i.e., $\mathbf s = {{\bf{A}}^H}{\bf{x}}$, where ${\bf{A}}={{\bf{\Lambda}} _{{c_2}}}{\bf{F}}{{\bf{\Lambda }}_{{c_1}}}$ is the DAFT matrix, ${\bf{F}} \in {\mathbb{C}^{N \times N}}$ is the normalized $N$-point discrete Fourier transform (DFT) matrix, $(\cdot)^H$ denotes the conjugate transpose, $c_1$ and $c_2$ are the AFDM chirp parameters, and
{\small
\begin{equation}
\label{deqn_ex1}
{{\bf{\Lambda}} _c} = {\rm{diag}}\left({e^{ - j2\pi c{n^2}}},\ n = 0,1, \ldots ,N - 1\right),\ c \in \{c_1, c_2\}.
\end{equation}
}

 Consider a doubly dispersive channel with $P$ propagation paths.  Let  $\tau_i$   and  $f_{i}$ denote the delay in seconds  and   Doppler frequency in hertz of the $i$-th path, respectively.  Then, the normalized delay is defined as ${l_i} = {l_i^{\rm{I}}} + {\iota _i} = \frac{{{\tau _i}}}{{{T_s}}}$, where ${l_i^{\rm{I}}}$ and ${\iota _i}$ are its integer and fractional parts, and ${{T_s}}$ is the sampling interval. Similarly, ${\nu _i} = {k_i^{\rm{I}}} + {\kappa _i} = N{f_i}{T_s}$ denotes the normalized Doppler shift, where ${k_i^{\rm{I}}}$ and ${\kappa _i}$ are its integer and fractional components, respectively. These definitions keep the delay and Doppler operators on a common normalized index.

\begin{figure}[tpb]
    \centering
    \subfloat[$l_i=0.1,\ {\nu _i}=0$]{
        \includegraphics[width=0.22\textwidth]{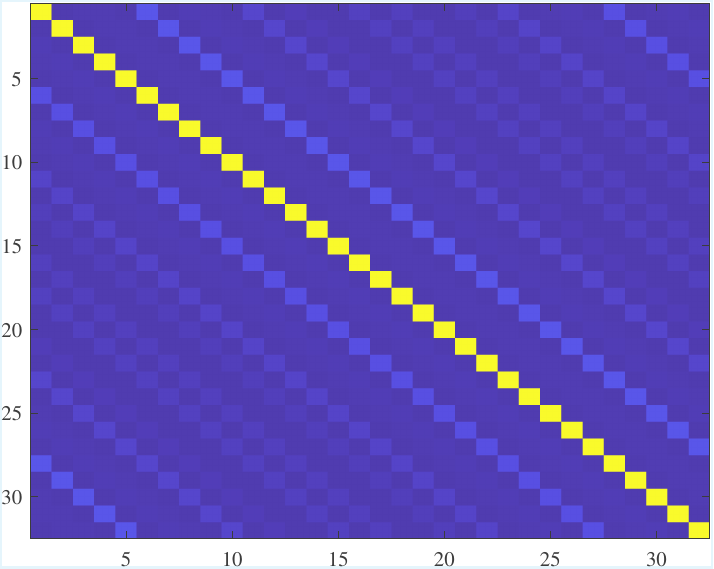}
        \label{fig:a}
    }
    \hfill
    \subfloat[$l_i=0.1,\ {\nu _i}=0.3$]{
        \includegraphics[width=0.22\textwidth]{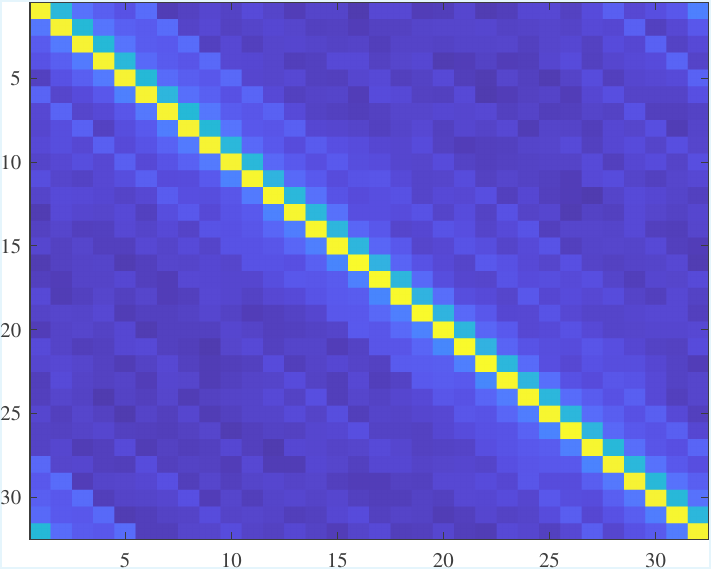}
        \label{fig:b}
    }
    \caption{ {Effects of fractional delay and Doppler on the DAFT-domain channel ($N=32$, $c_1= \frac{2}{N}$)}.}
    \label{fig_1}
\end{figure}

\color{black}
The input--output relationship derived in \cite{ref3,refXianle} assumes integer delays and therefore cannot accurately characterize FDFD channels. To circumvent this limitation, the fractional delay and Doppler shifts are equivalently represented as frequency-domain phase rotations and then mapped back to time-domain operators.  {We consider a narrowband system in which the carrier frequency
is much larger than the signal bandwidth and assume ideal sampling. Then, the    fractional delay of the $i$-th path  can be modeled by {${\bf{\Xi}}_i  = {{\bf{F}}^H}{\Delta _{{l_i}}}{\bf{F}}$} with ${\Delta _{{l_i}}} \buildrel \Delta \over = {\rm{diag}}({e^{ - j\frac{{2\pi {l_i}}}{N}n}},\ n = 0,1, \ldots ,N - 1)$ .
The received DAFT-domain signal under the FDFD channel is accordingly given by}
{\small
\begin{equation}
\begin{aligned}
\label{deqn_ex4}
{\bf{y}} = \sum\limits_{i = 1}^P {{h_i}\underbrace {{\bf{A}}{\Gamma _{{\rm{CP}}{{\rm{P}}_i}}} {{\Delta _{{\nu_i}}}}   {{\bf{\Xi}}_i}  {{\bf{A}}^H}}_{{\bf{H}}_i} {\bf{x}}}  + {\bf{Aw}} = {\bf{H}}_{{\rm{eff}}}{\bf{x}} + {\bf{\tilde w}}.
\end{aligned}
\end{equation}
}

\noindent where ${h_i}$ denotes the complex gain of the $i$-th path, ${\Gamma _{{\rm{CP}}{{\rm{P}}_i}}}$ is the chirp-periodic prefix (CPP) matrix \cite{ref3}, ${\bf{w}}$ is the additive white Gaussian noise (AWGN) vector, ${\bf{H}}_{{\rm{eff}}} = \sum\nolimits_{i=1}^{P} {h_i}{\bf{H}}_i$, and ${\bf{\tilde w}} = {\bf{Aw}}$.  {The diagonal matrix ${\Delta _{{\nu_i}}} \buildrel \Delta \over = {\rm{diag}}({e^{ - j2\pi {\nu_i} n / N}},\ n = 0,1, \ldots ,N - 1)$ models the Doppler of the $i$-th path.}  {In this paper, we assume \(c_1=\frac{C}{2N}\), with \(C\in\mathbb{Q}\) and even \(N\). Under these conditions, the CPP phase matrix becomes the identity matrix, and the CPP reduces to a conventional cyclic prefix (CP).}    The matrix  {${\bf{\Xi}}_i$} reduces to the conventional cyclic-shift matrix for integer delays, whereas for fractional delays it can be shown that
\begin{equation}
\begin{aligned}
\label{deqn_ex5}
{{\bf{\Xi}}_i} = \sum\limits_{k = 0}^{N - 1} {{{\rm{D}}_N}\left( {\frac{{k - {\iota _i}}}{N}} \right){{\bf{\Pi}} ^{k + {l_i^{\rm{I}}}}}},
\end{aligned}
\end{equation}
where ${\bf{\Pi}}$ denotes the forward cyclic-shift permutation matrix defined in \cite{ref4,ref5,ref6}.  {The term ${{\rm{D}}_N}\left( x \right) \buildrel \Delta \over = \frac{1}{N}\sum\nolimits_{n = 0}^{N - 1} {e^{j2\pi nx}}$ is the Dirichlet kernel, whose slowly decaying sidelobes govern the fractional-delay-induced energy leakage.} By substituting \eqref{deqn_ex5} into \eqref{deqn_ex4}, the elements of matrix ${\bf{H}}_i$ can be expressed as 
\begin{equation}
\begin{aligned}
\label{deqn_ex8}
&{\bf{H}}_i\left[ {p,q} \right] \\
&= \sum\limits_{k = 0}^{N - 1} {{{\rm{D}}_N}\left( {\frac{{k - {\iota _i}}}{N}} \right)\sum\limits_{n = 0}^{N - 1} {{{\left[ {\bf{A}} \right]}_{p,n}}{{e}^{ - {j2\pi }\frac{{{\nu _i}n}}{N}}}{{\left[ {{{\bf{A}}^H}} \right]}_{q,n - k - {l_i^{\rm{I}}}}}} } \\
& = \frac{1}{N}{e^{j2\pi {c_2}\left( {{q^2} - {p^2}} \right)}}\sum\limits_{k = 0}^{N - 1} {{\cal F}_i^k\left( q \right){\cal D}_i^k\left( {p,q} \right)},
\end{aligned}
\end{equation}
where   ${\cal F}_i^k\left( q \right) = {{\rm{D}}_N}\left( \frac{k - {\iota _i}}{N} \right){e^{j\frac{{2\pi}}{N}\left( {N{c_1}{{\left( {k + {l_i^{\rm{I}}}} \right)}^2} - q\left( {k + {l_i^{\rm{I}}}} \right)} \right)}}$ characterizes the fractional-delay spreading of the $i$-th path and ${\cal D}_i^k\left( {p,q} \right) = {{\rm{D}}_N}\big(\frac{q - p - {\nu _i} - 2N{c_1}\left( {k + {l_i^{\rm{I}}}} \right)}{N} \big)$ models the joint effect of the integer delay and the Doppler spreading. It is evident from \eqref{deqn_ex8} that each fractional-delay tap is associated with a Doppler spreading kernel centered at a different position, so that fractional delay and Doppler jointly induce a two-dimensional energy diffusion in the DAFT domain and severely weaken the channel sparsity.  {Fig.~\ref{fig_1} illustrates the DAFT-domain effective channel associated with the $i$-th path for $N=32$ under  $(l_i,\nu_i)=(0.1,0)$ and $(0.1,0.3)$.}

 \begin{figure} [tpb]
\centering
\includegraphics[width=3.0in]{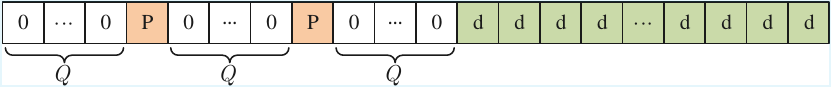}
\caption{Embedded dual-pilot symbol arrangement (‘P’: pilot, ‘d’: data, ‘0’: guard).}
\label{fig_2}
\vspace{-2em}
\end{figure}

\vspace{-0.5em}
\subsection{Channel Estimation Problem Formulation}
\vspace{-0.5em}

As shown in \eqref{deqn_ex8} and Fig.~\ref{fig_1}, the delay and Doppler parameters are inherently coupled in the FDFD channel. {To facilitate reliable channel estimation, an embedded dual-pilot structure is adopted, in which two pilot symbols are inserted at the $Q$-th and $(2Q+1)$-th DAFT-domain positions and the adjacent subcarriers are reserved as guard intervals to suppress coupled interference.    Since fractional delay and Doppler are coupled, we set the guard
length to
$Q \triangleq
\left\lceil
2Nc_1\left(l_{\max}+k_\tau\right)
+2\left(\nu_{\max}+k_\nu\right)
\right\rceil$, where $ \left\lceil \cdot  \right\rceil$ denotes the ceiling operator, $l_{\max}$ and $\nu_{\max}$ denote the maximum normalized
delay and Doppler shift, respectively, and $k_\tau$ and $k_\nu$
denote the additional guard margins introduced to account for fractional-delay and fractional-Doppler leakage, respectively.}  The resulting symbol arrangement is illustrated in Fig.~\ref{fig_2}.

{Under FDFD dispersion, each fractional-delay path response extends over the entire DAFT domain. To reduce the resulting interference from data symbols, channel estimation is performed using only $N_{\rm p}\leq Q$ received samples selected from each pilot observation window. Following \eqref{deqn_ex4}, the corresponding observation vector $\mathbf{y}_{\rm p}$ is given by}

\begin{equation}
\begin{aligned}
\label{deqn_ex10}
{{\bf{y}}_{\rm{p}}} = {{\bf{\Phi}} _{\rm{p}}}\left( {{\boldsymbol{l}},{\boldsymbol{\nu}} } \right){\bf{h}} + {{\bf{\tilde z}}_{\rm{p}}},
\end{aligned}
\end{equation}
where ${\boldsymbol{ l}} = {\left[ {{l_1},{l_2}, \ldots ,{l_P}} \right]^T} \in \mathbb{R}^{P \times 1}$ and ${\boldsymbol{\nu}}  = {\left[ {{\nu _1},{\nu _2}, \ldots ,{\nu _P}} \right]^T} \in \mathbb{R}^{P \times 1}$ denote the delay and Doppler vectors, respectively, and ${{\bf{\Phi}} _{\rm{p}}} \left( {{\boldsymbol{l}},{\boldsymbol{\nu}} } \right) = \left[ {{\boldsymbol{\varphi}_{\rm{p}}} \left( {{l_1},{\nu _1}} \right){\bf{x}}_{\rm{p}},{\boldsymbol{\varphi}_{\rm{p}}} \left( {{l_2},{\nu _2}} \right){\bf{x}}_{\rm{p}}, \ldots ,{\boldsymbol{\varphi}_{\rm{p}}} \left( {{l_P},{\nu _P}} \right){\bf{x}}_{\rm{p}}} \right] \in \mathbb{C} {^{N_{\rm{p}} \times P}}$ with ${\boldsymbol{\varphi} _{\rm{p}}}\left( {{l_i},{\nu _i}} \right) = {{\bf{A}}_{\rm{p}}}{\Gamma _{{\rm{CP}}{{\rm{P}}_i}}}{\Delta _{{\nu_i}}}{{\bf{F}}^H}{\Delta _{{l_i}}}{\bf{FA}}_{\rm{p}}^H$, where ${{\bf{A}}_{\rm{p}}}$ collects the rows of ${\bf{A}}$ associated with the retained samples and ${\bf{x}}_{\rm{p}}$ is the corresponding transmitted symbol vector. Moreover, ${\bf{h}} = {\left[ {{h_1},{h_2}, \ldots ,{h_{P}}} \right]^T}$ is the channel gain vector, and ${{\bf{\tilde z}}_{\rm{p}}}$ accounts for the AWGN and the residual data interference within the retained samples.
\vspace{-0.5em}

\section{Proposed SBL-Based Channel Estimation Algorithms}

{
In this section, we first develop the FLA-GR-SBL algorithm to address off-grid errors using a first-order approximation of the coupled delay--Doppler response. We then propose the 2D-GA-SBL algorithm to mitigate the associated approximation errors through iterative two-dimensional grid refinement.}

\vspace{-0.5em}
\subsection{Grid Refinement}
Since delay and Doppler parameters are continuous-valued, directly applying sparse recovery to \eqref{deqn_ex10} over a discrete grid suffers from off-grid model mismatch. Therefore, a grid-refinement strategy is adopted. Let ${N_l}$ and ${N_\nu }$ denote the numbers of points of a uniform virtual sampling grid in the delay and Doppler dimensions, with resolutions ${r_l} = \frac{l_{\max }}{N_l}$ and ${r_\nu } = \frac{2{\nu _{\max }}}{N_\nu}$, respectively. Casting channel estimation as a sparse signal recovery problem then yields
{\small
\begin{equation}
\begin{aligned}
\label{deqn_ex13}
{\bf{y}_{\rm p}} = {\bar {\bf{\Phi}}}_{\rm p}\left( {{\boldsymbol{\bar l}},\bar {\boldsymbol{\nu}} } \right){\bf{\bar h}} + {{\bf{\bar z}}_{\rm{p}}},
\end{aligned}
\end{equation}
}

\noindent where ${\boldsymbol{\bar l}} = {\left[ {{{\bar l}_0},{{\bar l}_1}, \ldots ,{{\bar l}_{{N_l}{N_\nu } - 1}}} \right]^T} \in \mathbb{R} {^{{N_l}{N_\nu } \times 1}}$, $\bar {\boldsymbol{\nu}}  = {\left[ {{{\bar \nu }_0},{{\bar \nu }_1}, \ldots ,{{\bar \nu }_{{N_l}{N_\nu } - 1}}} \right]^T} \in \mathbb{R} {^{{N_l}{N_\nu } \times 1}}$, and ${\bf{\bar h}} \in \mathbb{C} {^{{N_l}{N_\nu } \times 1}}$ denote the delay, Doppler, and channel gain components after virtual grid sampling, respectively. ${\bar {\bf{\Phi}}}_{\rm{p}} = \left[ {{{\bar {\boldsymbol{\phi}}_{\rm{p}} }}\left( {{{\bar l}_0},{{\bar \nu }_0}} \right),{{\bar {\boldsymbol{\phi}}_{\rm{p}} }}\left( {{{\bar l}_1},{{\bar \nu }_1}} \right), \ldots ,{{\bar {\boldsymbol{\phi}}_{\rm{p}} }}\left( {{{\bar l}_{{N_l}{N_\nu } - 1}},{{\bar \nu }_{{N_l}{N_\nu } - 1}}} \right)} \right] \in \mathbb{C} {^{{N}_{\rm p} \times {N_l}{N_\nu }}}$ with ${\bar {\boldsymbol{\phi}}_{\rm{p}} }\left( {{{\bar l}_{\bar i}},{{\bar \nu }_{\bar i}}} \right) = {\bar {\boldsymbol{\varphi}}_{\rm{p}}}\left( {{{\bar l}_{\bar i}},{{\bar \nu }_{\bar i}}} \right){{\bf{x}}_{\rm{p}}}$, and ${{\bf{\bar z}}_{\rm{p}}}$ additionally incorporates the grid-mismatch error.

\vspace{-1em}

\subsection{First-Order Linear Approximation}
In \eqref{deqn_ex13}, the true delay and Doppler parameters remain coupled and generally deviate from the virtual grid. To mitigate the resulting model mismatch, each dictionary column is linearized around its nearest grid point via the first-order approximation as
\begin{equation}
\begin{aligned}
 \small
\label{deqn_ex16}
{\bar {\boldsymbol{\phi}} _{\rm{p}}}\left( {{l_i},{\nu _i}} \right) \approx {\bar {\boldsymbol{\phi}} _{\rm{p}}}\left( {{{\bar l}_{\bar i}},{{\bar \nu }_{\bar i}}} \right) + {\bar {\boldsymbol{\phi}} '_{{\rm{p,}}l}}\left( {{{\bar l}_{\bar i}},{{\bar \nu }_{\bar i}}} \right){{\iota _{\bar i}}} + {\bar {\boldsymbol{\phi}} '_{{\rm{p,}}\nu }}\left( {{{\bar l}_{\bar i}},{{\bar \nu }_{\bar i}}} \right){\kappa _{\bar i}},
\end{aligned}
\end{equation}
  where ${\bar l_{\bar i}}$ and ${\bar \nu _{\bar i}}$ denote the grid points closest to ${l_i}$ and ${\nu_i}$, respectively, ${\bar {\boldsymbol{\phi}} '_{{\rm{p,}}l}}\left( {l,\nu } \right) = \frac{{\partial {{\bar {\boldsymbol{\phi}} }_{\rm{p}}}\left( {l,\nu } \right)}}{{\partial l}} \in \mathbb{C} {^{{N_{\rm{p}}} \times 1}}$ and ${\bar {\boldsymbol{\phi}} '_{{\rm{p,}}\nu}}\left( {l,\nu } \right) = \frac{{\partial {{\bar {\boldsymbol{\phi}} }_{\rm{p}}}\left( {l,\nu } \right)}}{{\partial \nu}} \in \mathbb{C} {^{{N_{\rm{p}}} \times 1}}$ are the corresponding first-order partial derivatives, and ${\iota _{\bar i}} = {l_i} - {{\bar l}_{\bar i}} \in \left[ { - \frac{1}{2}{r_l},\frac{1}{2}{r_l}} \right]$ and ${\kappa _{\bar i}} = {\nu _i} - {{\bar \nu }_{\bar i}} \in \left[ { - \frac{1}{2}{r_\nu },\frac{1}{2}{r_\nu }} \right]$ denote the off-grid delay and Doppler offsets of the $i$-th path, respectively. Substituting \eqref{deqn_ex16} into \eqref{deqn_ex13} gives
{\small
\begin{equation}
\begin{aligned}
\label{deqn_ex17}
{{\bf{y}}_{\rm{p}}} = \bar {\boldsymbol{\Phi}} _{\rm{p}}^{{\rm{1D}}}\left( {{{\boldsymbol{\iota}} _{\rm{v}}},{{\boldsymbol{\kappa}} _{\rm{v}}}} \right){{\bf{\bar h}}^{{\rm{1D}}}} + {\bf{\bar z}}_{\rm{p}}^{{\rm{1D}}},
\end{aligned}
\end{equation}
}

\noindent where the new noise vector ${\bf{\bar z}}_{\rm{p}}^{{\rm{1D}}}$ incorporates the linear approximation error, data interference, and white Gaussian  noise. The updated measurement matrix $\bar {\boldsymbol{\Phi}} _{\rm{p}}^{{\rm{1D}}}\left( {{{\boldsymbol{\iota}} _{\rm{v}}},{{\boldsymbol{\kappa}} _{\rm{v}}}} \right)$ can be expressed as
{\small
\begin{equation}
\begin{aligned}
\label{deqn_ex18}
\bar {\boldsymbol{\Phi}} _{\rm{p}}^{{\rm{1D}}}\left( {{{\boldsymbol{\iota}} _{\rm{v}}},{{\boldsymbol{\kappa}} _{\rm{v}}}} \right) = {\bar {\boldsymbol{\Phi}} _{\rm{p}}} + {\bar {\boldsymbol{\Phi}} _{{\rm{p,}}l}}{\rm{diag}}\left( {{{\boldsymbol{\iota}} _{\rm{v}}}} \right) + {\bar {\boldsymbol{\Phi}} _{{\rm{p,}}\nu }}{\rm{diag}}\left( {{{\boldsymbol{\kappa}} _{\rm{v}}}} \right),
\end{aligned}
\end{equation}
}

\noindent where ${\bar {\boldsymbol{\Phi}} _{{\rm{p}},l}},{\bar {\boldsymbol{\Phi}} _{{\rm{p}},\nu}} \in  \mathbb{C} {^{{N_{\rm{p}}} \times {N_l}{N_\nu }}}$ { collect the partial derivatives} ${\bar {\boldsymbol{\phi}} '_{{\rm{p,}}l}}$ and ${\bar {\boldsymbol{\phi}} '_{{\rm{p,}}\nu}}$ evaluated on the virtual grid $\left( {{\boldsymbol{\bar l}},{\boldsymbol{\bar \nu}}} \right)$, and ${{\boldsymbol{\iota}} _{\rm{v}}},{{\boldsymbol{\kappa}} _{\rm{v}}} \in  \mathbb{R} {^{{N_l}{N_\nu } \times 1}}$ collect the corresponding off-grid offsets.

In \eqref{deqn_ex17}, the channel gain vector ${\mathbf{\bar h}}^{\mathrm{1D}}$ and the off-grid parameter vectors ${\boldsymbol{\iota}}_{\mathrm{v}}$ and ${\boldsymbol{\kappa}}_{\mathrm{v}}$ constitute the three unknown quantities to be jointly estimated. Since only the entries corresponding to the $P$ propagation paths are nonzero, these vectors are all $P$-sparse, with the on-grid components determined by the support of ${\mathbf{\bar h}}^{\mathrm{1D}}$ and the off-grid components characterized by the estimated parameters ${\boldsymbol{\iota}}_{\mathrm{v}}$ and ${\boldsymbol{\kappa}}_{\mathrm{v}}$.

%In \eqref{deqn_ex17}, the vectors ${\mathbf{\bar h}}^{\mathrm{1D}}$, ${\boldsymbol{\iota}}_{\mathrm{v}}$, and ${\boldsymbol{\kappa}}_{\mathrm{v}}$ are the three unknown quantities to be jointly estimated. Since all other entries of ${\mathbf{\bar h}}^{\mathrm{1D}}$, ${\boldsymbol{\iota}}_{\mathrm{v}}$, and ${\boldsymbol{\kappa}}_{\mathrm{v}}$ are zero except those corresponding to the propagation paths, these vectors are all $P$-sparse. The on-grid components are determined by the support of ${\mathbf{\bar h}}^{\mathrm{1D}}$, while the off-grid components are characterized by the estimated parameters ${\boldsymbol{\iota}}_{\mathrm{v}}$ and ${\boldsymbol{\kappa}}_{\mathrm{v}}$.

\subsection{FLA-GR-SBL Channel Estimation}

This subsection develops the FLA-GR-SBL algorithm to estimate the unknown parameters in \eqref{deqn_ex17}. By embedding the first-order linear approximation into the grid-refinement framework \cite{ref6}, the algorithm effectively mitigates the off-grid errors caused by the delay--Doppler coupling of FDFD channels.

The likelihood function corresponding to \eqref{deqn_ex17} can be expressed as
{\small
\begin{equation}
\begin{aligned}
\label{deqn_ex20}
p\left( {\left. {{{\bf{y}}_{\rm{p}}}} \right|{{{\bf{\bar h}}}^{{\rm{1D}}}},\rho } \right) = {\left( {\frac{\rho }{\pi }} \right)^{{N_{\rm{p}}}}}{e^{ - \rho {{\left\| {{{\bf{y}}_{\rm{p}}} - \bar {\boldsymbol{\Phi}} _{\rm{p}}^{{\rm{1D}}}{{{\bf{\bar h}}}^{{\rm{1D}}}}} \right\|}^2}}},
\end{aligned}
\end{equation}
}

%To prevent overfitting, ${{\bf{\bar h}}^{{\rm{1D}}}}$ is assumed to follow a zero-mean complex Gaussian prior distribution with covariance matrix ${\rm{diag}}{\left( {\boldsymbol{\theta}}  \right)^{ - 1}}$. Accordingly, the probability density function of its $\bar i$-th element can be expressed as
\noindent where $\rho$ denotes the noise precision. Following the SBL framework, ${{\bf{\bar h}}^{{\rm{1D}}}}$ is assigned a zero-mean complex Gaussian prior with covariance matrix ${\rm{diag}}{\left( {\boldsymbol{\theta}}  \right)^{ - 1}}$, i.e.,
{\small
\begin{equation}
\begin{aligned}
\label{deqn_ex21}
p\left( {\left. {{{{\bf{\bar h}}}^{{\rm{1D}}}}} \right|{\boldsymbol{\theta}} } \right) = {\pi ^{ - {N_l}{N_\nu }}}\Big( {\prod\limits_{\bar i = 0}^{{N_l}{N_\nu } - 1} {\theta_ {\bar i}} } \Big) {e^{ - {{\left( {{{{\bf{\bar h}}}^{{\rm{1D}}}}} \right)}^H}{\rm{diag}}\left( {\boldsymbol{\theta}}  \right){{{\bf{\bar h}}}^{{\rm{1D}}}}}} ,
\end{aligned}
\end{equation}
}

\noindent where ${\boldsymbol{\theta}}  = {\left[ {\theta_0,\theta_1, \ldots ,\theta_{{N_l}{N_\nu } - 1}} \right]^T}$ is the precision hyperparameter vector. As conjugate priors of the Gaussian precisions, $\rho $ and ${\boldsymbol{\theta}} $ are further modeled by Gamma distributions, i.e.,
\begin{equation}
\small
\begin{aligned}
\label{deqn_ex22}
p\left( {\boldsymbol{\theta}}  \right) = \prod\limits_{\bar i = 0}^{{N_l}{N_\nu } - 1} {\Gamma \left( {{\theta _{\bar i}};a,b} \right)} ,\ p\left( \rho  \right) = \Gamma \left( {\rho ;c,d} \right),
\end{aligned}
\end{equation}

\noindent where $\Gamma \left(  \cdot\, ; \cdot , \cdot  \right)$ denotes the Gamma distribution with shape--rate hyperparameter pairs $\left( a,b \right)$ and $\left( c,d \right)$, respectively. The unknown off-grid components are modeled as uniformly distributed, i.e., ${\iota _{\bar i}} \sim {\cal U}\left[ { - \frac{1}{2}{r_l},\frac{1}{2}{r_l}} \right]$ and ${\kappa _{\bar i}} \sim {\cal U}\left[ { - \frac{1}{2}{r_\nu },\frac{1}{2}{r_\nu }} \right]$.

{To estimate $\bar{\mathbf{h}}^{\mathrm{1D}}$, we first obtain its posterior distribution, which can be expressed using Bayes' theorem as}
{\small
\begin{equation}
\begin{aligned}
\label{deqn_ex24}
p\left( \left. \bar{\mathbf{h}}^{\mathrm{1D}},\boldsymbol{\theta},
\boldsymbol{\iota}_{\mathrm{v}},\boldsymbol{\kappa}_{\mathrm{v}},\rho \right| \mathbf{y}_{\mathrm{p}} \right)
&= p\left( \left. \bar{\mathbf{h}}^{\mathrm{1D}} \right|
\mathbf{y}_{\mathrm{p}},\boldsymbol{\theta},
\boldsymbol{\iota}_{\mathrm{v}},\boldsymbol{\kappa}_{\mathrm{v}},\rho \right) \\
&\quad \times
p\left( \left. \boldsymbol{\theta},
\boldsymbol{\iota}_{\mathrm{v}},\boldsymbol{\kappa}_{\mathrm{v}},\rho
\right| \mathbf{y}_{\mathrm{p}} \right).
\end{aligned}
\end{equation}
}

Therefore, at the $t$-th iteration, given the hyperparameters $\left( {{{\boldsymbol{\theta}} ^{\left( t \right)}},{\boldsymbol{\iota}} _{\rm{v}}^{\left( t \right)},{\boldsymbol{\kappa}} _{\rm{v}}^{\left( t \right)},{\rho ^{\left( t \right)}}} \right)$, the conditional posterior distribution of ${{\bf{\bar h}}^{{\rm{1D}}}}$ can be expressed as
{\small
\begin{equation}
\begin{aligned}
\label{deqn_ex26}
&p\left( \left. \bar{\mathbf{h}}^{\mathrm{1D}} \right|
\mathbf{y}_{\mathrm{p}};{{{\boldsymbol{\theta}} ^{\left( t \right)}},{\boldsymbol{\iota}} _{\rm{v}}^{\left( t \right)},{\boldsymbol{\kappa}} _{\rm{v}}^{\left( t \right)},{\rho ^{\left( t \right)}}} \right)\\
&= {\pi ^{ - {N_l}{N_\nu }}}{\left| {{\boldsymbol{\Sigma}} _{{\bf{\bar h}}}^{\left( t \right)}} \right|^{ - 1}} {e^{ - {{\left( {{{{\bf{\bar h}}}^{{\rm{1D}}}} - {\boldsymbol{\mu}} _{{\bf{\bar h}}}^{\left( t \right)}} \right)}^H}{{\left( {{\boldsymbol{\Sigma}} _{{\bf{\bar h}}}^{\left( t \right)}} \right)}^{ - 1}}\left( {{{{\bf{\bar h}}}^{{\rm{1D}}}} - {\boldsymbol{\mu}} _{{\bf{\bar h}}}^{\left( t \right)}} \right)}},
\end{aligned}
\end{equation}
}

\noindent where the posterior covariance matrix and the posterior mean are respectively given by

\begin{equation}
\begin{aligned}
\small
\label{deqn_ex27}
\begin{array}{c}
{\bf{\Sigma}} _{{\bf{\bar h}}}^{\left( t \right)} = {\left( {{\rho ^{\left( t \right)}}{{\left( {\bar {\bf{\Phi}} _{\rm{p}}^{{\rm{1D}}}} \right)}^H}\bar {\bf{\Phi}} _{\rm{p}}^{{\rm{1D}}} + {\rm{diag}}\left( {{{\boldsymbol{\theta}} ^{\left( t \right)}}} \right)} \right)^{ - 1}},\\
{\boldsymbol{\mu}} _{{\bf{\bar h}}}^{\left( t \right)} = {\rho ^{\left( t \right)}}{\bf{\Sigma}} _{{\bf{\bar h}}}^{\left( t \right)}{\left( {\bar {\bf{\Phi}} _{\rm{p}}^{{\rm{1D}}}} \right)^H}{{\bf{y}}_{\rm{p}}}.
\end{array}
\end{aligned}
\end{equation}
The hyperparameters $\left( {{\boldsymbol{\theta}} ,{{\boldsymbol{\iota}} _{\rm{v}}},{{\boldsymbol{\kappa}} _{\rm{v}}},\rho } \right)$ are then updated by maximizing the posterior $p\left( {\left. {{\boldsymbol{\theta}} ,{{\boldsymbol{\iota}} _{\rm{v}}},{{\boldsymbol{\kappa}} _{\rm{v}}},\rho } \right|{{\bf{y}}_{\rm{p}}}} \right) \propto p\left( {{{\bf{y}}_{\rm{p}}},{{\boldsymbol{\theta}} ,{{\boldsymbol{\iota}} _{\rm{v}}},{{\boldsymbol{\kappa}} _{\rm{v}}},\rho } } \right)$, where the joint distribution factorizes as
\begin{equation}
\small
\begin{aligned}
\label{deqn_ex28}
&p\left( {{{\bf{y}}_{\rm{p}}},{{{\bf{\bar h}}}^{{\rm{1D}}}},{{\boldsymbol{\theta}} ,{{\boldsymbol{\iota}} _{\rm{v}}},{{\boldsymbol{\kappa}} _{\rm{v}}},\rho } } \right)\\
& = p\left( {\left. {{{\bf{y}}_{\rm{p}}}} \right|{{{\bf{\bar h}}}^{{\rm{1D}}}};{{{\boldsymbol{\iota}} _{\rm{v}}},{{\boldsymbol{\kappa}} _{\rm{v}}},\rho } } \right) p\left( {  {{{{\bf{\bar h}}}^{{\rm{1D}}}}}  | {\boldsymbol{\theta}} } \right)p\left( {\boldsymbol{\theta}}  \right)p\left( {{{\boldsymbol{\iota}} _{\rm{v}}}} \right)p\left( {{{\boldsymbol{\kappa}} _{\rm{v}}}} \right)p\left( \rho  \right).
\end{aligned}
\end{equation}
The expectation-maximization (EM) algorithm \cite{ref13} is adopted to update the hyperparameters at the $\left( t+1 \right)$-th iteration. In the expectation step, the expectation of the log joint distribution in \eqref{deqn_ex28} is taken with respect to the conditional posterior of ${{\bf{\bar h}}^{{\rm{1D}}}}$, i.e.,
{\small
\begin{equation}
\begin{aligned}
\label{deqn_ex29}
&Q\left( {\left. {{\boldsymbol{\theta}} ,{{\boldsymbol{\iota}} _{\rm{v}}},{{\boldsymbol{\kappa}} _{\rm{v}}},\rho } \right|{{{\boldsymbol{\theta}} ^{\left( t \right)}},{\boldsymbol{\iota}} _{\rm{v}}^{\left( t \right)},{\boldsymbol{\kappa}} _{\rm{v}}^{\left( t \right)},{\rho ^{\left( t \right)}}}} \right)\\
& = {\mathbb{E}_{\left( {\left. {{{{\bf{\bar h}}}^{{\rm{1D}}}}} \right|{{\bf{y}}_{\rm{p}}};{{{\boldsymbol{\theta}} ^{\left( t \right)}},{\boldsymbol{\iota}} _{\rm{v}}^{\left( t \right)},{\boldsymbol{\kappa}} _{\rm{v}}^{\left( t \right)},{\rho ^{\left( t \right)}}}} \right)}}\left\{ {\ln \left[ {p\left( {{{\bf{y}}_{\rm{p}}},{{{\bf{\bar h}}}^{{\rm{1D}}}},{{\boldsymbol{\theta}} ,{{\boldsymbol{\iota}} _{\rm{v}}},{{\boldsymbol{\kappa}} _{\rm{v}}},\rho } } \right)} \right]} \right\}.
\end{aligned}
\end{equation}
}

In the maximization step, each hyperparameter is optimized alternately. Setting the derivative of \eqref{deqn_ex29} with respect to ${\theta _{\bar i}}$ to zero yields
{\small
\begin{equation}
\begin{aligned}
\label{deqn_ex35}
\theta _{\bar i}^{\left( {t + 1} \right)} = \frac{a}{{b + {\bf{\Sigma}} _{{\bf{\bar h}}}^{\left( t \right)}\left[ {\bar i,\bar i} \right] + {{\left| {{\boldsymbol{\mu}} _{{\bf{\bar h}}}^{\left( t \right)}\left[ {\bar i} \right]} \right|}^2}}}, \ \bar i = 0,1, \ldots ,{N_l}{N_\nu } - 1,
\end{aligned}
\end{equation}
}

\noindent where ${\bf{\Sigma}} _{{\bf{\bar h}}}^{\left( t \right)}\left[ {\bar i,\bar i} \right]$ and ${\boldsymbol{\mu}} _{{\bf{\bar h}}}^{\left( t \right)}\left[ {\bar i} \right]$ denote the $\left( {\bar i,\bar i} \right)$-th element of ${\bf{\Sigma}} _{{\bf{\bar h}}}^{\left( t \right)}$ and the $\bar i$-th element of ${\boldsymbol{\mu}} _{{\bf{\bar h}}}^{\left( t \right)}$ in \eqref{deqn_ex27}, respectively. Similarly, the noise precision is updated as
{\small
\begin{equation}
\begin{aligned}
\label{deqn_ex36}
{\rho ^{\left( {t + 1} \right)}} = ({{c + {N_{\rm{p}}} - 1}})/ {\tilde d},
\end{aligned}
\end{equation}
}

\noindent where $\tilde d = d + {\left\| {{{\bf{y}}_{\rm{p}}} - \bar {\bf{\Phi}} _{\rm{p}}^{{\rm{1D}}}{\boldsymbol{\mu}} _{{\bf{\bar h}}}^{\left( t \right)}} \right\|^2} + {\rm{Tr}}\left( {\bar {\bf{\Phi}} _{\rm{p}}^{{\rm{1D}}}{\bf{\Sigma}} _{{\bf{\bar h}}}^{\left( t \right)}{{\left( {\bar {\bf{\Phi}} _{\rm{p}}^{{\rm{1D}}}} \right)}^H}} \right)$ and ${\rm{Tr}}\left(  \cdot  \right)$ denotes the matrix trace.

Note that although ${{\bf{\bar h}}^{{\rm{1D}}}}$ in \eqref{deqn_ex17} contains ${N_l}{N_\nu }$ elements, only the entries associated with the $P$ propagation paths are significant. According to compressive sensing theory \cite{ref10}, up to $\bar P = \left\lfloor {\frac{{{N_{\rm{p}}}}}{{\ln \left( {{N_l}{N_\nu }} \right)}}} \right\rfloor$ nonzero elements can be reliably recovered from ${N_{\rm{p}}}$ measurements; hence, the number of paths is initialized as $\bar P$.

Maximizing \eqref{deqn_ex29} with respect to the off-grid delay components yields the element-wise update
\begin{equation}
\small
\begin{aligned}
\label{deqn_ex41}
{\left[ {{\boldsymbol{\iota}}_{{\rm{p}},l}^{\left( {t + 1} \right)}} \right]_n} = \frac{{{{\left[ {{{\bf{b}}_{{\rm{p}},l}}} \right]}_n} - \left[ {{{\left[ {{{\bf{A}}_{{\rm{p}},l}}} \right]}_n}} \right]_{ - n}^T{{\left[ {{{\boldsymbol{\iota}}_{{\rm{p}},l}}} \right]}_{ - n}}}}{{{{\left[ {{{\bf{A}}_{{\rm{p}},l}}} \right]}_{n,n}}}},
\end{aligned}
\end{equation}
 where ${[  \cdot  ]_n}$ denotes the $n$-th element of a vector, ${[  \cdot  ]_{-n}}$ denotes the vector with its $n$-th element removed, and ${[  \cdot  ]_{m,n}}$ denotes the $( {m,n} )$-th entry of a matrix. {Let $S_l^{\left( t \right)}$, with $| {S_l^{\left( t \right)}} | = \bar P$, denote the index set of the $\bar P$ smallest elements of ${{\boldsymbol{\theta}} ^{\left( t \right)}}$, i.e., the estimated path support. } Then, ${{\bf{A}}_{{\rm{p,}}l}} \in \mathbb{R} {^{\bar P \times \bar P}}$ and ${{\bf{b}}_{{\rm{p,}}l}} \in \mathbb{R} {^{\bar P \times 1}}$ are obtained by extracting from ${{\bf{A}}_{\boldsymbol{\iota}} }$ and ${{\bf{b}}_{\boldsymbol{\iota}} }$ the entries indexed by $S_l^{\left( t \right)}$, where ${{\bf{A}}_{\boldsymbol{\iota}} }$ and ${{\bf{b}}_{\boldsymbol{\iota}} }$ are given in \eqref{deqn_ex39} at the bottom of this page, with $\Re \left\{  \cdot  \right\}$ and $\odot$ denoting the real part and the Hadamard product, respectively.

\setcounter{equation}{20}
\begin{figure*}[hb]
\hrulefill
{\small
\begin{equation}
\begin{aligned}
\label{deqn_ex39}
\begin{array}{c}
{{\bf{A}}_{\boldsymbol{\iota}} } = \Re \left\{ {{{\left( {\bar {\bf{\Phi}} _{{\rm{p,}}l}^{{\rm{1D}}}} \right)}^H}\bar {\bf{\Phi}} _{{\rm{p,}}l}^{{\rm{1D}}} \odot \left( {{{\left( {{\boldsymbol{\mu}} _{{\bf{\bar h}}}^{\left( t \right)}} \right)}^*}{{\left( {{\boldsymbol{\mu}} _{{\bf{\bar h}}}^{\left( t \right)}} \right)}^T} + {{\left( {{\bf{\Sigma}} _{{\bf{\bar h}}}^{\left( t \right)}} \right)}^T}} \right)} \right\},\\
{{\bf{b}}_\iota } = \Re \left\{ {{\rm{diag}}\left( {{\boldsymbol{\mu}} _{{\bf{\bar h}}}^{\left( t \right)}} \right){{\left( {\bar {\bf{\Phi}} _{{\rm{p,}}l}^{{\rm{1D}}}} \right)}^T}{\bf{y}}_{\rm{p}}^* - {\rm{diag}}\left( {\left( {{\boldsymbol{\mu}} _{{\bf{\bar h}}}^{\left( t \right)}{{\left( {{\boldsymbol{\mu}} _{{\bf{\bar h}}}^{\left( t \right)}} \right)}^H} + {\bf{\Sigma}} _{{\bf{\bar h}}}^{\left( t \right)}} \right){{\left( {\bar {\bf{\Phi}} _{\rm{p}}^{{\rm{1D}}} + \bar {\bf{\Phi}} _{{\rm{p,}}\nu }^{{\rm{1D}}}{\rm{diag}}\left( {{\boldsymbol{\kappa}} _{\rm{v}}^{\left( t \right)}} \right)} \right)}^H}\bar {\bf{\Phi}} _{{\rm{p,}}l}^{{\rm{1D}}}} \right)} \right\}.
\end{array}
\end{aligned}
\end{equation}
}

\vspace{-0.5em}
\setcounter{equation}{24}
{\small
\begin{equation}
\begin{aligned}
\label{deqn_ex44}
\begin{array}{c}
{{\bf{A}}_{\boldsymbol{\kappa}} } = \Re \left\{ {{{\left( {\bar {\bf{\Phi}} _{{\rm{p,}}\nu }^{{\rm{1D}}}} \right)}^H}\bar {\bf{\Phi}} _{{\rm{p,}}\nu }^{{\rm{1D}}} \odot \left( {{{\left( {{\boldsymbol{\mu}} _{{\bf{\bar h}}}^{\left( t \right)}} \right)}^*}{{\left( {{\boldsymbol{\mu}} _{{\bf{\bar h}}}^{\left( t \right)}} \right)}^T} + {{\left( {{\bf{\Sigma}} _{{\bf{\bar h}}}^{\left( t \right)}} \right)}^T}} \right)} \right\},\\
{{\bf{b}}_{\boldsymbol{\kappa}} } = \Re \left\{ {{\rm{diag}}\left( {{\boldsymbol{\mu}} _{{\bf{\bar h}}}^{\left( t \right)}} \right){{\left( {\bar {\bf{\Phi}} _{{\rm{p,}}\nu }^{{\rm{1D}}}} \right)}^T}{\bf{y}}_{\rm{p}}^* - {\rm{diag}}\left( {\left( {{\boldsymbol{\mu}} _{{\bf{\bar h}}}^{\left( t \right)}{{\left( {{\boldsymbol{\mu}} _{{\bf{\bar h}}}^{\left( t \right)}} \right)}^H} + {\bf{\Sigma}} _{{\bf{\bar h}}}^{\left( t \right)}} \right){{\left( {\bar {\bf{\Phi}} _{\rm{p}}^{{\rm{1D}}} + \bar {\bf{\Phi}} _{{\rm{p,}}l}^{{\rm{1D}}}{\rm{diag}}\left( {{\boldsymbol{\iota}} _{\rm{v}}^{\left( t \right)}} \right)} \right)}^H}\bar {\bf{\Phi}}_{{\rm{p,}}\nu }^{{\rm{1D}}}} \right)} \right\}.
\end{array}
\end{aligned}
\end{equation}
}

\vspace{-0.5em}
\setcounter{equation}{28}
{\small
\begin{equation}
\begin{aligned}
\label{deqn_ex49}
{\bar {\boldsymbol{\phi}} _{\rm{p}}}\left( {l_i^{\left( t \right)},\nu _i^{\left( t \right)}} \right) \approx {\bar {\boldsymbol{\phi}} _{\rm{p}}}\left( {\bar l_{\bar i}^{\left( {t - 1} \right)},\bar \nu _{\bar i}^{\left( {t - 1} \right)}} \right) + {\bar {\boldsymbol{\phi}} '_{{\rm{p,}}l}}\left( {\bar l_{\bar i}^{\left( {t - 1} \right)},\bar \nu _{\bar i}^{\left( {t - 1} \right)}} \right)\Delta _{{\iota _{\bar i}}}^{\left( t \right)} + {\bar {\boldsymbol{\phi}} '_{{\rm{p,}}\nu }}\left( {\bar l_{\bar i}^{\left( {t - 1} \right)},\bar \nu _{\bar i}^{\left( {t - 1} \right)}} \right)\Delta _{{\kappa _{\bar i}}}^{\left( t \right)},
\end{aligned}
\end{equation}
}

\end{figure*}

When ${{\bf{A}}_{{\rm{p,}}l}}$ is invertible, the update has the following closed-form expression
\setcounter{equation}{21}

\begin{equation}
\small
\begin{aligned}
\label{deqn_ex40}
{{\boldsymbol{\iota}} _{{\rm{p,}}l}^{\left( {t + 1} \right)}} = {\left( {{{\bf{A}}_{{\rm{p,}}l}}} \right)^{ - 1}}{{\bf{b}}_{{\rm{p,}}l}}.
\end{aligned}
\end{equation}
Since ${\iota _{\bar i}} \in \left[ { - \frac{{{r_l}}}{2},\frac{{{r_l}}}{2}} \right]$, the estimate is clipped as
\begin{equation}
\begin{aligned}
\small
\label{deqn_ex42}
{\left[ {\overline {{\boldsymbol{\iota}}}_{{\rm{p}},l}^{\left( {t + 1} \right)} } \right]_n} = \left\{ {\begin{array}{*{20}{l}}
{{{\left[ {{\boldsymbol{\iota}}_{{\rm{p}},l}^{\left( {t + 1} \right)}} \right]}_n},}&{{{\left[ {{\boldsymbol{\iota}}_{{\rm{p}},l}^{\left( {t + 1} \right)}} \right]}_n} \in \left[ { - \frac{1}{2}{r_l},\frac{1}{2}{r_l}} \right],}\\
{ - \frac{1}{2}{r_l},}&{{{\left[ {{\boldsymbol{\iota}}_{{\rm{p}},l}^{\left( {t + 1} \right)}} \right]}_n} <  - \frac{1}{2}{r_l},}\\
{\frac{1}{2}{r_l},}&{{\rm{otherwise}}.}
\end{array}} \right.
\end{aligned}
\end{equation}
The off-grid delay estimate ${\boldsymbol{\iota}} _{\rm{v}}^{\left( {t + 1} \right)}$ is then obtained by placing $\overline {{\boldsymbol{\iota}}} _{{\rm{p,}}l}^{\left( {t + 1} \right)}  $ from \eqref{deqn_ex42} onto the support $S_l^{\left( t \right)}$.

Similarly, the off-grid Doppler component ${{\boldsymbol{\kappa}} _{\rm{v}}}$ is updated element-wise as
{\small
\begin{equation}
\begin{aligned}
\label{deqn_ex46}
{\left[ {{\boldsymbol{\kappa}} _{{\rm{p}},\nu }^{\left( {t + 1} \right)}} \right]_n} = \frac{{{{\left[ {{{\bf{b}}_{{\rm{p}},\nu }}} \right]}_n} - \left[ {{{\left[ {{{\bf{A}}_{{\rm{p}},\nu }}} \right]}_n}} \right]_{ - n}^T{{\left[ {{{\boldsymbol{\kappa}} _{{\rm{p}},\nu }}} \right]}_{ - n}}}}{{{{\left[ {{{\bf{A}}_{{\rm{p}},\nu }}} \right]}_{n,n}}}}.
\end{aligned}
\end{equation}
}

\noindent where ${{\bf{A}}_{{\rm{p,}}\nu }} \in \mathbb{R} {^{\bar P \times \bar P}}$ and ${{\bf{b}}_{{\rm{p,}}\nu }} \in \mathbb{R} {^{\bar P \times 1}}$ are extracted, in the same manner, from ${{\bf{A}}_{\boldsymbol{\kappa}} }$ and ${{\bf{b}}_{\boldsymbol{\kappa}} }$ given in \eqref{deqn_ex44} at the bottom of this page. If $\mathbf{A}_{\mathrm{p},\nu}$ is invertible, the update has the following closed-form expression:
\setcounter{equation}{25}
{\small
\begin{equation}
\begin{aligned}
\label{deqn_ex45}
{{\boldsymbol{\kappa}} _{{\rm{p,}}\nu }^{\left( {t + 1} \right)}} =  {{\bf{A}}_{{\rm{p,}}\nu }^{ - 1}}{{\bf{b}}_{{\rm{p,}}\nu }}.
\end{aligned}
\end{equation}
}

Since ${\kappa _{\bar i}} \in \left[ { - \frac{{{r_\nu}}}{2},\frac{{{r_\nu}}}{2}} \right]$, the estimate is clipped as
{\small
\begin{equation}
\begin{aligned}
\label{deqn_ex47}
{\left[ {\overline {{\boldsymbol{\kappa}}} _{{\rm{p}},\nu }^{\left( {t + 1} \right)} } \right]_n} = \left\{ {\begin{array}{*{20}{l}}
{{{\left[ {{\boldsymbol{\kappa}} _{{\rm{p}},\nu }^{\left( {t + 1} \right)}} \right]}_n},}&{{{\left[ {{\boldsymbol{\kappa}} _{{\rm{p}},\nu }^{\left( {t + 1} \right)}} \right]}_n} \in \left[ { - \frac{1}{2}{r_\nu },\frac{1}{2}{r_\nu }} \right],}\\
{ - \frac{1}{2}{r_\nu },}&{{{\left[ {{\boldsymbol{\kappa}} _{{\rm{p}},\nu }^{\left( {t + 1} \right)}} \right]}_n} <  - \frac{1}{2}{r_\nu },}\\
{\frac{1}{2}{r_\nu },}&{{\rm{otherwise}}.}
\end{array}} \right.
\end{aligned}
\end{equation}
}

Analogously, ${{\boldsymbol{\kappa}} _{\rm{v}}^{\left( {t + 1} \right)}}$ is obtained by placing $\overline {{\boldsymbol{\kappa}} }_{{\rm{p,}}\nu }^{\left( {t + 1} \right)} $ from \eqref{deqn_ex47} onto the support $S_\nu ^{\left( t \right)}$.

In summary, the channel gain is updated through the mean ${\boldsymbol{\mu}} _{{\bf{\bar h}}}^{\left( t \right)}$ of the channel vector ${{\bf{\bar h}}^{{\rm{1D}}}}$ in \eqref{deqn_ex27}. Meanwhile, the hyperparameters, as well as the estimates of delay and Doppler shift, are iteratively updated according to \eqref{deqn_ex35}, \eqref{deqn_ex36}, \eqref{deqn_ex42}, and \eqref{deqn_ex47}. The iteration is terminated when the maximum number of iterations $t = {T_{\max }}$ is reached or when the convergence condition $\frac{{{{\left\| {{{\boldsymbol{\theta}} ^{\left( {t + 1} \right)}} - {{\boldsymbol{\theta}} ^{\left( t \right)}}} \right\|}_2}}}{{{{\left\| {{{\boldsymbol{\theta}} ^{\left( t \right)}}} \right\|}_2}}} < \xi $ is satisfied. Finally, the estimated delay and Doppler  shifts are given by ${\bf{\hat l}} = {{\boldsymbol{\bar l}}_{\rm{v}}} + {\boldsymbol{\iota}} _{\rm{v}}^{\left( t \right)}$ and $\hat {\boldsymbol{\nu}}  = {\bar {\boldsymbol{\nu}} _{\rm{v}}} + {\boldsymbol{\kappa}} _{\rm{v}}^{\left( t \right)}$, respectively, and the channel gain on the virtual sampling grid is given by $\hat{\bar{\mathbf{h}}}^{\mathrm{1D}} = \boldsymbol{\mu}_{\bar{\mathbf{h}}}^{(t)}$.

%\vspace{-1em}
\subsection{2D-GA-SBL Channel Estimation}

In FLA-GR-SBL, when the true delay or Doppler lies   between adjacent grid points, the refined sensing matrix exhibits strong column correlation, yielding an ill-conditioned Gram matrix. In addition, together with the limited accuracy of the first-order approximation, this may cause the off-grid estimates to shrink toward zero or oscillate without converging. To address these issues, the 2D-GA-SBL method is proposed, which iteratively refines the two-dimensional delay--Doppler grid to improve both stability and accuracy.

To this end, the first-order approximation model in \eqref{deqn_ex16} is reformulated as
\begin{equation}
\small
\begin{aligned}
\label{deqn_ex48}
{\bar {\boldsymbol{\phi}} _{\rm{p}}}\left( {{l_i},{\nu _i}} \right) \approx {\bar {\boldsymbol{\phi}} _{\rm{p}}}\left( {{{\bar l}_{\bar i}},{{\bar \nu }_{\bar i}}} \right) + {\bar {\boldsymbol{\phi}} '_{{\rm{p,}}l}}\left( {{{\bar l}_{\bar i}},{{\bar \nu }_{\bar i}}} \right){\Delta _{{\iota _{\bar i}}}} + {\bar {\boldsymbol{\phi}} '_{{\rm{p,}}\nu }}\left( {{{\bar l}_{\bar i}},{{\bar \nu }_{\bar i}}} \right){\Delta _{{\kappa _{\bar i}}}},
\end{aligned}
\end{equation}
where ${\Delta _{{\iota _{\bar i}}}} = {l_i} - {\bar l_{\bar i}}$ and ${\Delta _{{\kappa _{\bar i}}}} = {\nu _i} - {\bar \nu _{\bar i}}$ denote the delay and Doppler approximation increments, respectively. At the $t$-th iteration, the approximation is performed around the grid points refined in the previous iteration, as given in \eqref{deqn_ex49} at the bottom of this page.

In \eqref{deqn_ex49},  $\Delta _{{\iota _{\bar i}}}^{\left( t \right)}$ and $\Delta _{{\kappa _{\bar i}}}^{\left( t \right)}$ are obtained from \eqref{deqn_ex42} and \eqref{deqn_ex47}, respectively, corresponding to the estimates of ${\boldsymbol{\iota}} _{\rm{v}}^{\left( t \right)}$ and ${\boldsymbol{\kappa}} _{\rm{v}}^{\left( t \right)}$. Meanwhile, $\Delta _{{\iota _{\bar i}}}^{\left( t \right)}$ and $\Delta _{{\kappa _{\bar i}}}^{\left( t \right)}$ are required to satisfy
\setcounter{equation}{29}
\begin{equation}
\small
\begin{aligned}
\label{deqn_ex50}
\left| {\Delta _{{\iota _{\bar i}}}^{\left( t \right)} + \chi _{{\iota _{\bar i}}}^{\left( {t - 1} \right)}} \right| < \frac{{{l_{\max }}}}{{{N_l}}}, \qquad  {\rm{ }}\left| {\Delta _{{\kappa _{\bar i}}}^{\left( t \right)} + \chi _{{\kappa _{\bar i}}}^{\left( {t - 1} \right)}} \right| < \frac{{2{\nu _{\max }}}}{{{N_\nu }}},
\end{aligned}
\end{equation}
where $\chi _{{\iota _{\bar i}}}^{\left( {t - 1} \right)} = \Delta _{{\iota _{\bar i}}}^{\left( 1 \right)} + \Delta _{{\iota _{\bar i}}}^{\left( 2 \right)} +  \cdots  + \Delta _{{\iota _{\bar i}}}^{\left( {t - 1} \right)}$ and $\chi _{{\kappa _{\bar i}}}^{\left( {t - 1} \right)} = \Delta _{{\kappa _{\bar i}}}^{\left( 1 \right)} + \Delta _{{\kappa _{\bar i}}}^{\left( 2 \right)} +  \cdots  + \Delta _{{\kappa _{\bar i}}}^{\left( {t - 1} \right)}$ denote the accumulated delay and Doppler shift approximation components over the first $\left(t - 1\right)$ iterations, respectively. 

Through this iterative refinement, the grid points progressively converge toward the true delay and Doppler values, thereby avoiding the error floor and instability of the one-shot first-order approximation.

 {
Since both algorithms update the posterior of all $N_lN_\nu$
candidate coefficients, they share the same posterior-update
complexity. Specifically, constructing and factorizing the
posterior precision matrix in \eqref{deqn_ex27} has a dominant computational
complexity of
$\mathcal O\big(N_{\rm p}(N_lN_\nu)^2+(N_lN_\nu)^3\big)$
per iteration. The difference in computational cost lies in the   dictionary
updating. Specifically, FLA-GR-SBL precomputes the base dictionary and derivative
matrices in \eqref{deqn_ex18}, whereas 2D-GA-SBL reevaluates the
dictionary columns and their two derivatives at the $\bar P$
relocated grid points for the local approximation in
\eqref{deqn_ex49}. By implementing the DFT operations in the pilot
dictionary defined after \eqref{deqn_ex10} using FFTs, these
additional evaluations incur a complexity of
$\mathcal O(\bar P N\log N)$ per iteration.
The updated dictionary is then used in the common posterior
update in \eqref{deqn_ex27}, without requiring a separate matrix
inversion for each relocated point.
}

\begin{figure*}[!t]
\centering

\begin{minipage}[t]{0.24\textwidth}
    \centering
    \includegraphics[width=\linewidth]{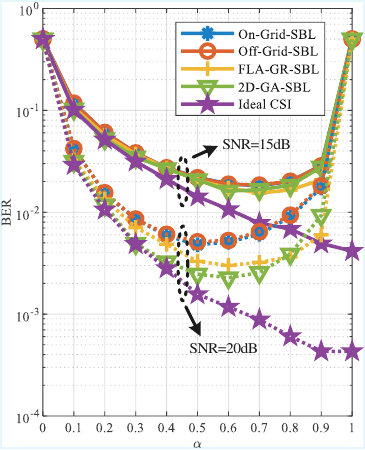}
    \caption{BER versus the power allocation factor $\alpha$.}
    \label{fig_3}
\end{minipage}
\hfill
\begin{minipage}[t]{0.24\textwidth}
    \centering
    \includegraphics[width=\linewidth]{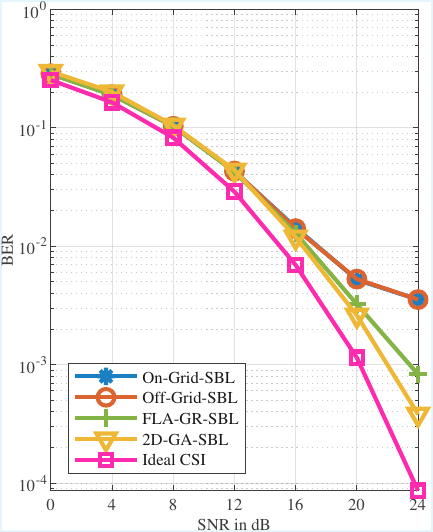}
    \caption{BER versus SNR of different channel estimation algorithms ($\alpha=0.6$).}
    \label{fig_4}
\end{minipage}
\hfill
\begin{minipage}[t]{0.24\textwidth}
    \centering
    \includegraphics[width=\linewidth]{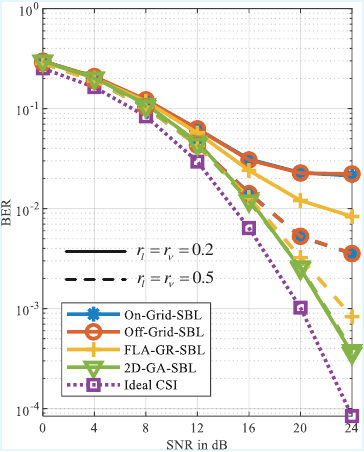}
    \caption{BER under different grid resolutions $r_l$ and $r_\nu$.}
    \label{fig_5}
\end{minipage}
\hfill
\begin{minipage}[t]{0.24\textwidth}
    \centering
    \includegraphics[width=\linewidth]{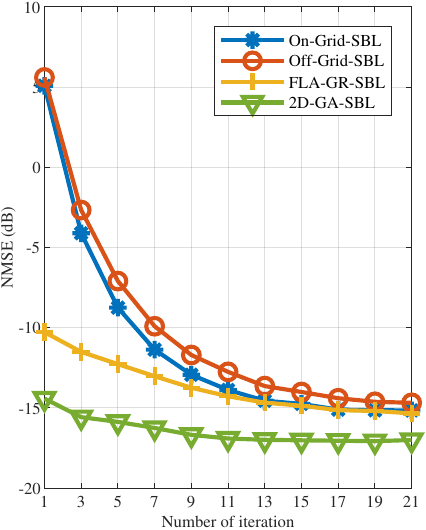}
    \caption{NMSE versus the number of iterations (${r_l}={r_\nu}=0.5$, ${\rm{SNR}}=20$ dB).}
    \label{fig_7}
\end{minipage}
\vspace{-2em}
\end{figure*}
\section{Simulation Results}
This section evaluates the proposed algorithms, with the on-grid and off-grid SBL methods of \cite{ref10} as benchmarks. The number of AFDM subcarriers is $N = 256$, the carrier frequency is ${f_{\rm{c}}} = 24$~GHz, and the subcarrier spacing is $f_{\rm{AFDM}} = 15$~kHz. We consider a $3$-path channel with a maximum normalized delay of $l_{\max}=4$, where the normalized delay of each path is independently drawn from a continuous uniform distribution over $[0,l_{\max}]$.  The maximum normalized Doppler shift is defined as ${\nu _{\max }} = \frac{{{f_{\rm{c}}} v}}{{c f_{\rm{AFDM}}}}$, where $v$ and $c$ denote the terminal velocity and the speed of light, respectively. We set ${\nu _{\max }}=1$, corresponding to $v=675$ km/h. The Jakes Doppler spectrum is adopted, i.e., the normalized Doppler shift of the $i$-th path is ${\nu _i} = {\nu _{\max }}\cos ({\psi _i})$, where ${\psi _i}$ is uniformly distributed over $[ - \pi ,\pi ]$. The pilot and data signal-to-noise ratios (SNRs) are defined as ${\rm{SN}}{{\rm{R}}_{\rm{p}}} = 10\log_{10}({\left| {{x_{\rm{p}}}} \right|^2}/{\sigma ^2})$ and ${\rm{SN}}{{\rm{R}}_{\rm{d}}} = 10\log_{10} (\mathbb{E}\{ {| {{x_{\rm{d}}}}|^2}\} /{\sigma ^2})$. {Unless otherwise specified, we set $c_1=\frac{3}{2N}$, the guard parameters to $k_\tau=1$ and $k_\nu=1$, and use $Q=19$ and $N_{\mathrm{P}}=16$.}

{
 A power allocation factor
$\alpha\in[0,1]$ specifies the fraction assigned to data
transmission, yielding total data and pilot powers of
$\alpha P_{\rm s}$ and $(1-\alpha)P_{\rm s}$, respectively.
 For $l_{\max}=4$ and $\nu_{\max}=1$, the grid spacings
$r_l=r_\nu=0.5$ and $r_l=r_\nu=0.2$ correspond to
$(N_l,N_\nu)=(8,4)$ and $(20,10)$, respectively.
The finer grid, $r_l=r_\nu=0.2$, is used in
Figs.~\ref{fig_3} and~\ref{fig_4}.
For both proposed algorithms, the precision vector is
initialized as $\boldsymbol{\theta}^{(0)}
=\mathbf{1}_{N_lN_\nu}$, and the off-grid delay and Doppler
offsets are initialized to zero.
The maximum number of iterations and the convergence
tolerance are set to $T_{\max}=50$ and $\xi=10^{-3}$,
respectively.}

Fig.~\ref{fig_3} shows the BER of different channel estimation algorithms versus the power allocation factor $\alpha$.  As can be seen from the figure, the BER first decreases and then increases with $\alpha$, attaining its minimum around $\alpha =0.6$, which shows the trade-off between channel estimation and data detection. The proposed FLA-GR-SBL and 2D-GA-SBL algorithms achieve a notable improvement over the conventional on-grid and off-grid SBL methods.
In particular, the proposed 2D-GA-SBL performs the best over the whole range of $\alpha$ by accurately capturing the delay--Doppler coupling and suppressing energy leakage.

Fig.~\ref{fig_4} further compares the BER versus  ${\rm{SN}}{{\rm{R}}_{\rm{d}}}$ at $\alpha =0.6$.  The on-grid and off-grid SBL methods remain performance limited
over the entire SNR range, as they fail to characterize
the FDFD-induced energy spreading.  The proposed FLA-GR-SBL improves the performance by mitigating off-grid errors through the first-order approximation and grid refinement, while 2D-GA-SBL performs the best since its two-dimensional grid approximation captures the delay--Doppler coupling more accurately.

Fig.~\ref{fig_5} compares the BER performance for two grid resolutions, \(r_l=r_\nu=0.2\) and \(r_l=r_\nu=0.5\). The on-grid and off-grid SBL schemes exhibit BER degradation as the grid is refined, indicating sensitivity to grid resolution. In contrast, the proposed FLA-GR-SBL outperforms these baselines under both settings, whereas the proposed 2D-GA-SBL consistently achieves the lowest BER with little variation between the two grid resolutions, demonstrating its robustness for FDFD channels.

 Finally, Fig.~\ref{fig_7} compares the NMSE convergence behavior of the considered algorithms. All algorithms converge in approximately \(16\) iterations. The on-grid and off-grid SBL baselines exhibit relatively high NMSE values, whereas FLA-GR-SBL achieves a lower NMSE by mitigating off-grid modeling errors through grid refinement. The proposed 2D-GA-SBL exhibits stable convergence and attains the lowest steady-state NMSE among all evaluated schemes.

\vspace{-1em}

\section{Conclusion}
This paper investigated channel estimation for AFDM systems over doubly dispersive channels with fractional delay and fractional Doppler shifts. We first formulated the corresponding DAFT-domain input--output relationship and characterized the resulting coupled delay--Doppler spreading. Based on this model, the FLA-GR-SBL algorithm was developed to jointly estimate the channel gains and the off-grid delay and Doppler components using a first-order approximation. To reduce the modeling error associated with fixed expansion points, the 2D-GA-SBL algorithm was further developed to iteratively relocate the grid points in both the delay and Doppler dimensions. Simulation results under the considered channel and system settings showed that the proposed algorithms achieve lower NMSE and BER than the conventional on-grid and off-grid SBL benchmarks, with 2D-GA-SBL providing more stable performance across the tested grid resolutions. Future work will investigate robust guard-interval design, colored residual-interference modeling, unknown path-number estimation, and low-complexity implementations for more practical vehicular channel models.

\end{document}